\documentclass[useAMS,usenatbib]{mn2e}
\usepackage{graphicx}% Include figure files
\usepackage{times}% Include figure files
\newcommand{\sss}{\scriptstyle}

\def\vtp{v_{{\sss T}p}}
\def\vte{v_{{\sss T}e}}

\def\vta{v_{{\sss T}a}}

\def\vt{v_{{\sss T}}}
\newcommand {\be}{\begin{equation}} % start equation
\newcommand{\ee}{\end{equation}}    % end equation

\title[Viscosity effects on waves  in  partially and fully ionized plasma in magnetic field ]{Viscosity effects on waves  in  partially and fully ionized plasma in magnetic field }
\author[J. Vranjes]{J. Vranjes$^{1}$\thanks{E-mail:
jvranjes@yahoo.com}\\
$^{1}$Institute of Physics Belgrade, Pregrevica 118, 11080 Zemun, Serbia.\\
}

\begin{document}

\date{Accepted x. Received y; in original form z}

\pagerange{\pageref{firstpage}--\pageref{lastpage}} \pubyear{2002}

\maketitle

\label{firstpage}

\begin{abstract}
 Viscosity is discussed in  multicomponent partially and fully ionized plasma, and its  effects on  two very different waves (Alfv\'{e}n and Langmuir) in solar atmosphere. A  full set of viscosity coefficients is presented which includes coefficients
 for electrons,  protons and hydrogen atoms. These are applied to layers with mostly magnetized protons in solar chromosphere where the Alfv\'{e}n wave
 could   in principle  be expected. The viscosity coefficients are calculated and presented graphically for the altitudes between 700 and 2200 km, and required corresponding cross sections for various types of collisions are given in terms of altitude. It is shown that in chromosphere the viscosity plays no role for the Alfv\'{e}n wave, which is only strongly affected by ion friction with neutrals. In corona, assuming the magnetic  field of a few Gauss, the Alfv\'{e}n wave is more affected by  ion viscosity than by ion-electron friction only for wavelengths shorter that 1-30 km, dependent on parameters and assuming  the perturbed magnetic field of one percent of its equilibrium value. For the Langmuir wave the viscosity-friction interplay in chromosphere is shown to be dependent on altitude and on wavelengths. In corona the viscosity is the main dissipative mechanism acting on the Langmuir mode.

\end{abstract}

\begin{keywords}
Plasmas; Waves; Sun: photosphere - chromosphere - corona - fundamental parameters
\end{keywords}

\section{Introduction }\label{s1}

In fully ionized plasmas, or in plasmas with predominant Coulomb collisions, the ion viscosity coefficients are relatively simple in the limits $\Omega_i\gg \nu_i$ or $\Omega_i\ll \nu_i$. Here, $\Omega_i, \nu_i$ are the ion gyro-frequency and collision frequency, and the limits given here generally describe different ion magnetization regimes.  The former implies a magnetized plasma where both kinematic and gyro-viscosity may play a role, and the latter implies a plasma where the magnetic field plays no practical role. The celebrated work of \citet{bra}, and his text in the book   \citet{bra2} have been  basic references in this field for many decades.
The first  deals with a fully ionized plasma, and the second also includes chapters where neutrals are taken into account. However, as it may be seen from the literature ever since, it is sometimes  overlooked that in \citet{bra2}  the viscosity in chapters dealing with neutrals is {\em explicitly omitted}. In the literature, his  expressions for viscosity derived in his texts for purely electron-ion plasma have been inappropriately  used also in studies which include neutrals. This may not be justified in weakly ionized plasmas where the amount of neutrals exceeds the amount of ions for several orders of magnitude, like in the solar photosphere and lower chromosphere.

The general method used in Braginskii's  works for calculation of viscosity components and other transport coefficients is based on the Chapman-Enskog method, which  implies an expansion of the distribution function with an underlying necessity that terms in the expansion converge sufficiently rapidly, if at all. Braginskii himself writes about some uncertainty in satisfying this crucial condition. Some later works [e.g. \citet{epe}] indicate some  inaccuracies  in Braginskii's thermoelectric and heat flow coefficients caused by the truncation errors incorporated in the method, and these inaccuracies depend on the magnetization regime.

These issues should be kept in mind when using transport coefficients from various available sources. The most urgent for the solar plasma seems to be need for reliable and clear expressions for viscosity coefficients in partly ionized plasma in which ions are magnetized, like in the upper solar chromosphere. Such expressions are greatly needed for both analytical and numerical studies in view of an increased interest of researchers in phenomena in weakly ionized solar atmosphere. Rather general and complicated expressions may be found in the literature, like those in the book of \citet{zh},  and in \citet{sh} and \citet{sn}.
For un-magnetized photosphere and lower chromosphere all relevant transport coefficients have been presented recently in \citet{vkr}, calculated from the starting BGK collisional model integral. On the other hand, for plasma in magnetic field which affects particle motion, the most complete and detailed theory is presented in \citet{zh}. The theory is based on the Grad's method (i.e., Hermitian moment method) which dates back to 1949 \citep{gr}. Both Braginskii and Zhdanov results agree remarkably well, although different methods are used. However, the theory presented in \citet{zh} is definitely more systematic and extensive, and  it includes general multicomponent  and multitemperature plasmas  with an arbitrary number of charged species together with neutrals, as well as polyatomic gas mixtures, which are missing in the Braginskii's theory.

This Zhdanov's theory will be used in the present work in application to Alfv\'{e}n wave propagation in solar atmosphere. In our recent paper \citet{vko} the effects of collisions on the  Alfv\'{e}n wave in photosphere and lower chromosphere were studied using the most accurate collision cross sections for proton-hydrogen collisions, which include several essential features intrinsic to such a weakly ionized environment, obtained in \citet{vkr}. These include the charge exchange, the quantum-mechanical effect of indistinguishability for colliding protons and hydrogen atoms at energies below 1 eV, and the effects of polarization of neutral atoms in the process of collisions with charged protons. The results from \citet{vko} show non-existence of Alfv\'{e}n waves in the lower solar atmosphere in a layer that is at least about 600 km wide (assuming very strong magnetic structures with magnetic field of $B_0=0.1$ T constant with altitude), and possibly even two times wider in a more realistic case of the same starting field which is  decreasing  with altitude. In such a layer protons are not magnetized and this is the reason why Alfv\'{e}n waves (AW) cannot be excited. All relevant transport coefficients for viscosity and thermal conductivity have been derived in \citet{vkr}.
Though, viscosity effects have been omitted in \citet{vko}  in order to have our full multi-component model as close as possible to the classic MHD works, in order to explore similarities and differences between the two models. The analysis was restricted to the lower layers where the AW is shown to be either non-existent or heavily damped by friction, so additional damping due to viscosity was not essential in any case.

However, physical parameters change with  altitude and the ratio $\Omega_i/\nu_i$  changes as well, being much less than unity in the photosphere \citep{vkr}, and above unity in the upper chromosphere. This means that above  certain altitudes  the AW  may be expected to propagate although as a strongly damped mode, and viscosity effects may play a role and should be included in such a way as to be able to follow this altitude dependent variation of parameters. Kinematic viscosity (associated  mainly with  neutrals) is predominant in lower layers (photosphere), this kinematic viscosity (of both ions and neutrals)  is then accompanied with gyro-viscosity in chromosphere and in neighboring  upper layers.

The aim of this work is twofold, i) to provide a reliable and self-consistent set of expressions for viscosity in any plasma (partially or totally ionized), and ii) to apply these results to the plasma in solar atmosphere in order to check the role of viscosity on the propagation of some waves, and for this purpose we have focused on two very different ones,  Alfv\'{e}n and Langmuir waves. Both kinematic dissipative viscosity and gyro-viscosity coefficients are presented in Sec.~\ref{s2} for partially ionized plasma, and for fully ionized plasma in Sec.~\ref{s3}. In Sec.~\ref{s4} the effects of dissipations are presented in detail for the case of AW in chromosphere, and in particular the relative importance of viscosity (in comparison to friction) is discussed for both waves in the chromosphere and corona.

\section{Partially ionized plasma }\label{s2}

\subsection{Viscosity coefficients for ions and neutrals }\label{s2a}

Using the notation from \citet{zh}, the ion viscous stress tensor is of the shape:
\begin{eqnarray}
\Pi_{irs} &= &-\eta_{i0} W_{rs 0}- \eta_{i1} W_{rs 1} -\eta_{i2} W_{rs 2} +\eta_{i3} W_{rs 3} \nonumber \\
  & & +\eta_{i4} W_{rs 4}, \quad r\in x, y, z, \quad s \in x, y, z,
\label{e1}
\end{eqnarray}
where  $W_{rsj}$, $j\in (0, \cdots 4)$  are various contractions  \citep{bra2} of the traceless rate-of-strain tensor  $W_{rs}$ (here and further in the text with suppressed index for the species) which is given through a general coordinate $\vec \zeta$ with components $r$ or $s$ as:
\be
W_{rs}=\frac{\partial v_j}{\partial \zeta_k} + \frac{\partial v_k}{\partial \zeta_j}- \frac{2}{3} \delta_{jk} \nabla\cdot \vec v, \quad \zeta_j, \zeta_k \in x, y, z. \label{e1a}
\ee
The components of $W_{rs}$ are:
\[
W_{xx}=\frac{4}{3}\frac{\partial v_x}{\partial x} - \frac{2}{3} \left(\frac{\partial v_y}{\partial y} +  \frac{\partial v_z}{\partial z}\right),
\]
\[
W_{yy}=\frac{4}{3}\frac{\partial v_y}{\partial y} - \frac{2}{3} \left(\frac{\partial v_x}{\partial x} +  \frac{\partial v_z}{\partial z}\right),
\]
\[
W_{zz}=\frac{4}{3}\frac{\partial v_z}{\partial z} - \frac{2}{3} \left(\frac{\partial v_x}{\partial x} +  \frac{\partial v_y}{\partial y}\right),
\]
\[
W_{xy}=\frac{\partial v_x}{\partial y} + \frac{\partial v_y}{\partial x}=W_{yx}, \quad W_{xz}=\frac{\partial v_x}{\partial z} + \frac{\partial v_z}{\partial x}=W_{zx},
\]
\[
W_{yz}=\frac{\partial v_y}{\partial z} + \frac{\partial v_z}{\partial y}=W_{zy}.
\]
 Without loss of generality  the magnetic field may be assumed oriented in the $x$-direction, and this yields  the following components of the ion stress tensor $\Pi_{irs}$:
\begin{eqnarray}
\Pi_{ixx} &=& -\eta_{i0}W_{xx}, \nonumber\\
\Pi_{iyy} &=& -\frac{\eta_{i0}}{2}\left(W_{yy} + W_{zz}\right)-\frac{\eta_{i1}}{2}\left(W_{yy} - W_{zz}\right) -\eta_{i3}W_{yz},  \nonumber\\
\Pi_{izz} &=& -\frac{\eta_{i0}}{2}\left(W_{yy} + W_{zz}\right)-\frac{\eta_{i1}}{2}\left(W_{zz} - W_{yy}\right) +\eta_{i3}W_{yz},  \nonumber\\
\Pi_{iyz} &=& \Pi_{izy}=-\eta_{i1}W_{yz}+ \frac{\eta_{i3}}{2}\left(W_{yy}-W_{zz}\right),  \nonumber\\
\Pi_{ixy} &=& \Pi_{iyx}=-\eta_{i2}W_{xy} -\eta_{i4}W_{xz},  \nonumber\\
\Pi_{ixz} &=& \Pi_{izx}=-\eta_{i2}W_{xz} +\eta_{i4}W_{xy}. \label{st}
 \end{eqnarray}
Further we shall use  $m_i=m_a=m$, $T_i=T_a=T$. In this case the viscosity coefficients for ions are \citep{zh}:
\[
\eta_{i0}=\frac{p_i \tau_i \xi_i \Delta_{\eta}^{-1}}{2}, \quad \eta_{i1}=\frac{\eta_{i0}}{1+b_i^2 \Delta_{\eta}^{-2}}, \quad \eta_{i3} =\eta_{i1} b_i \Delta_{\eta}^{-1},
\]
\[
\eta_{i2}=\eta_{i1}\left[\frac{b_i}{2}\right]=\frac{\eta_{i0}}{1+\frac{\displaystyle{b_i^2 \Delta_{\eta}^{-2}}}{\displaystyle{4}}},
\]
\be
\eta_{i4}=\eta_{i3}\left[\frac{ b_i}{2}\right]=\frac{b_i}{2} \frac{\eta_{i0}  \Delta_{\eta}^{-1}}{1+ \frac{ \displaystyle{b_i^2  \Delta_{\eta}^{-2}}}{\displaystyle{4}}}.\label{vi}
\ee
For neutrals $a$ the shape of  $\Pi_{ars}$ is the same, and coefficients are:
\[
 \eta_{a0}=\frac{p_a \tau_a \xi_a \Delta_{\eta}^{-1}}{2}, \!\quad \eta_{a1}=\eta_{a0} \frac{1+b_i^2  \xi_a^{-1}\Delta_{\eta}^{-1}}{1+b_i^2 \Delta_{\eta}^{-2}},
\]
\[
\eta_{a2}=\eta_{a1}\left[\frac{b_i}{2}\right]=\eta_{a0} \frac{1+ \frac{\displaystyle{b_i^2}}{\displaystyle{4}}  \xi_a^{-1}\Delta_{\eta}^{-1}}{1+
\frac{\displaystyle{b_i^2}}{\displaystyle{4}} \Delta_{\eta}^{-2}},
\]
\[
\eta_{a3} =\eta_{a0} b_i \Delta_{\eta}^{-1}\, \frac{1- \xi_a^{-1} \Delta_{\eta}}{1+ b_i^2\Delta_{\eta}^{-2}},
\]
\be
 \eta_{a4}=\eta_{a3} \left[\frac{b_i}{2}\right]=\eta_{a0} \frac{\displaystyle{b_i}}{\displaystyle{2}} \Delta_{\eta}^{-1}\, \frac{1- \xi_a^{-1} \Delta_{\eta}}{1+ \frac{\displaystyle{b_i^2}}{\displaystyle{4}}\Delta_{\eta}^{-2}}. \label{nv}
\ee
All coefficients $\eta_{\alpha j}$ presented in Eqs.~(\ref{vi}, \ref{nv}) are valid within the limit $m_e/m\ll T/T_e$, which is easily satisfied in many plasmas. Various parameters that appear here are  \citep{zh}:
\[
p_j=n_j \kappa T_j, \quad b_i=\Omega_i \tau_i, \! \quad \Omega_i= \frac{e z_i B_0}{m}, \! \quad \xi_i=1+ f_{ia} \tau_a \tau_{ia}^{-1},
\]
\[
\xi_a=1+ f_{ia} \tau_i \tau_{ai}^{-1}, \quad \tau_i^{-1}=0.3 \tau_{ii}^{-1} + f_{ia}' \tau_{ia}^{-1} + \delta \tau_{ie}^{-1},
 \]
\[
\tau_a^{-1}=0.3 A_{aa}^*\tau_{aa}^{-1} + f_{ia}' \tau_{ai}^{-1}  + \delta \tau_{ae}^{-1}, \quad \delta= \frac{m_e}{m_j}, \quad j\neq e,
\]
\[
f_{ia}'=\frac{1}{4}\left(1+ 0.6 A_{ia}^*\right), \quad f_{ia}= \frac{1}{4}\left(1- 0.6 A_{ia}^*\right),
\]
\be
\Delta_{\eta}=1- a_{ia} a_{ai}, \quad a_{ia}=- \frac{f_{ia} \tau_i}{\tau_{ai}}, \quad a_{ai}=- \frac{f_{ia} \tau_a}{\tau_{ia}}. \label{par}
\ee
For collisions between hard spheres, and for Coulomb collisions $A_{\alpha \beta}^*=1$, so this value will be used,  but instead the usual hard sphere
radius of the hydrogen atom and protons, we shall use the most accurate  viscosity collision cross sections given in \citet{vkr}.

The complete sets of coefficients (\ref{vi}) and (\ref{nv}) clearly include both the usual kinematic viscosity associated with friction  and the collisionless gyro-viscosity.  They are very general, valid for any ratio $\Omega_i/\nu_i$, and can easily be reduced to various limiting cases regarding this ratio. It is seen that the two sets are  mutually coupled and as a result of this coupling the dynamics of neutrals is in principle affected by the ion gyro-effects as well.

Regarding the remaining parameters in the expressions (\ref{par}), we continue with the ion-ion collisional time  as given by \citet{zh}:
\be
\tau_{ii}=\frac{6 \varepsilon_0^2 m_i^{1/2} (\pi \kappa T_i)^{3/2}}{e^4 z_i^4 n_i L_{ii}}.\label{e2}
\ee
Note that in \cite{bra2}, this ion collision time  is taken greater by a factor 2. According to \citet{zh} this is because in Braginskii' book the relaxation  time is taken differently so that Braginskii's time is $\tau_b= 2 \tau_{ii}$. But in \citet{sh} the expression for $\tau_{ii}$ in fact coincides with $\tau_b$.  On the other hand, in  \citet{shk} and in \citet{hel} on the right-hand side in  Eq.~(\ref{e2}) one can find an additional numerical factor $2^{1/2}$,  whose origin is rather unclear.  We shall use the given value  (\ref{e2}) for $\tau_{ii}$; further in the text a perfect agreement  will be shown  between the results of Zhdanov and Braginskii bearing in mind this only difference by factor 2 in collisional time.

For Coulomb collisions between electrons and ions we shall use \citep{zh, hel}:
\be
\tau_{ei}=\frac{6 (2m_e)^{1/2} \varepsilon_0^2 (\pi \kappa T_e)^{3/2}}{ e^4 z^2 n_i L_{ei}}, \label{e3}
\ee
 and for ion-electron time the momentum conservation then yields:
 \[
 \tau_{ie}=\frac{m_i n_i}{m_e n_e} \tau_{ei}.
 \]
 Note that Eq.~(\ref{e3}) is the same as electron collision time $\tau_e$ given in \citet{bra2} where its meaning seems  to be the same, i.e., it implies e-i collisions.
 It is  useful to remember that $\tau_{ei}= \sqrt{2}\, \tau_{ee}$ [c.f.,  \cite{mk} and  \citet{zh}]. Both expressions $\tau_{ei}$ and $\tau_{ee}$ may also be obtained from a more general one given  in \citet{zh}:
 \[
 \tau_{\alpha \beta}^{-1}=\frac{n_{\beta} 16 \pi^{1/2}}{3}\left(\frac{\gamma_{\alpha \beta}}{2}\right)^{3/2} \left(\frac{q_{\alpha} q_{\beta}}{4 \pi \varepsilon_0 \mu_{\alpha \beta}}\right)^2 L_{\alpha \beta}.
 \]
   The Coulomb logarithm and other quantities used here read \citep{zh}:
 \[
 L_{\alpha \beta}=\ln \Lambda_{\alpha\beta}=\ln\left[\frac{12 \pi \varepsilon_0}{|q_{\alpha} q_{\beta}|} \frac{\mu_{\alpha\beta} r_d}{\gamma_{\alpha\beta}}\right], \quad \mu_{\alpha\beta}=\frac{m_{\alpha} m_{\beta}}{m_{\alpha}+ m_{\beta}},
 \]
\[
 \gamma_{\alpha\beta}=\frac{\gamma_{\alpha} \gamma_{\beta}}{\gamma_{\alpha} +\gamma_{\beta}}, \quad \gamma_j=\frac{m_j}{\kappa T_j}, \quad \frac{1}{r_d^2}=\sum_j\frac{n_jq_j^2}{\varepsilon_0 \kappa T_j}.
 \]
In case of many ion species $j$, instead of Eq.~(\ref{e3}) the electron collision time with all ions  becomes \citep{zh}:
\be
\tau_{ei}=\frac{6 (2m_e)^{1/2} \varepsilon_0^2 (\pi \kappa T_e)^{3/2}}{ e^4 \sum_j z_j^2 n_j L_{ej}}. \label{e4}
\ee
In this case the argument of the Coulomb logarithm in Eq.~(\ref{e4}) can be approximated (for $T_e=T_j=T$) by:
\[
\Lambda_{ej}\equiv \Lambda_1=\frac{12 \pi \kappa T \varepsilon_0^{3/2}}{  z_{ef}e^2 } \left[\frac{\kappa T}{n_e (1+ z_{ef})e^2}\right]^{1/2},
\]
\[
 z_{ef}=\frac{\sum_j n_j z_j^2}{\sum_j n_j z_j}=\frac{\sum_j n_j z_j^2}{n_e}.
 \]
On the other hand, in the case $T_e\gg T_j=T$, the Coulomb logarithm is:
 \[
\Lambda_{ej}\equiv \Lambda_2=\frac{12 \pi \kappa T_e \varepsilon_0^{3/2}}{ z_{ef} e^2} \left[\frac{\kappa T}{n_e z_{ef}e^2}\right]^{1/2}.
\]
 In electron-proton solar plasma with $T_e=T_i$ we shall thus use:
 \be
\Lambda_{ep}=\frac{2^{1/2}6\pi(\varepsilon_0 \kappa T)^{3/2}}{e^3n^{1/2}}. \label{cl}
\ee
Further, in equations (\ref{par}), for  $\tau_{ia}=1/\nu_{ia}$ we use p-H collision frequency $\nu_{p\tiny H}= \sigma_{p\tiny  H}n_{\tiny H} \vtp$ from \citet{vkr}, where for  $\sigma_{p \tiny H}$
 we have to use the viscosity line  4 from Fig.~1 in the same reference.

 For atom-atom (that is H-H)  collisions we shall use the integral viscosity cross section for quantum-mechanically indistinguishable nuclei given  by line 2 in Fig.~3 from \citet{vkr},  $\nu_{{\sss HH}}= \sigma_{{\sss HH}} n_{{\sss H}} v_{{\sss TH}}$. As shown in  \citet{vkr}, the dynamic viscosity coefficient obtained in this exact way is for about factor 2 smaller than the value obtained from the classic hard sphere modeling, and it is in very good agreement with experimental values.
 From now on we shall use the index $a$  instead of $H$, $T_a=T_i=T$, $v_{{\sss TH}}=\vta= \vtp=\vt$, and  $\nu_{{\sss HH}}=\nu_{aa}=1/\tau_{aa}$,  $\nu_{p\tiny H}= \nu_{ia}=1/\tau_{ia}$, etc.
  Momentum conservation further yields $\tau_{ai}=\tau_{ia} m_a n_a/(m_i n_i)$.

  For e-a collisions we have $\tau_{ae}=\tau_{ea} m_n n_n/(m_e n_e)$. For $\tau_{ea}$ we use Fig.~4 from \citet{vkr} which provides e-H cross section for elastic scattering, and Table~1 from \citet{vko} with values for momentum transfer. We stress that there is some uncertainty in the literature regarding measurement of electron cross section at low energies. On the other hand,  comparison reveals no practical differences between the two mentioned cross sections,  and we shall use the same value for the viscosity cross section as well.

 With all this we have completely determined numerous  parameters in Eq.~(\ref{par}) for partially ionized plasmas, and  what remains are only  particular plasma parameters (densities, temperature)  for various altitudes, which we shall use from  \citet{fon}. For the magnetic field we shall use some models with altitude dependent  magnitude \citep{le, vko}.

All cross sections for ion collisions with neutrals are dependent on relative energy of colliding particles. In the lower solar atmosphere the temperature changes with altitude and so does the energy of particles taking part in collisions. For this reason the cross sections are altitude dependent and it is useful to have them given at one place and ready for a direct use in this work or elsewhere. The temperature in \citet{fon} is given in K, in lab frame, and cross sections in \citet{vkr} are in eV in both lab frame and center of mass (CM) frame of colliding particles. So for the purpose of this work and in general it is convenient to present the cross sections with altitude, and in Table~\ref{t1} in Appendix~\ref{ap1} we give the altitude $h$ and the corresponding temperature from \citet{fon}  in  lab frame [which corresponds to top axis in Figs.~1,~2,~3 in \citet{vkr}], and the corresponding cross sections. Note that the  temperature (energy) in CM frame is $ T_{{\sss CM}} = T_{lab} m_2/(m_1+ m_2)$, where $m_{1,2}$ are masses of particles, so for proton-hydrogen collisions this implies that $T_{{\sss CM}} = T_{lab}/2$. The remaining columns in Table~\ref{t1} give cross sections for elastic scattering, momentum transfer and viscosity, for both $p-a$ and $a-a$ collisions. The only cross section which shows a (rather slight) monotonous decrease with altitude (i.e.., temperature)
is the viscosity cross section for atoms $\sigma_{aa,v}$. All others only show just the usual variations typical for low energies.  These data suggest that in the given altitude range $h\in 0, 2200$ km, it  may sometimes  be good enough to take the following approximate values for the cross sections for $p-a$ scattering, momentum transfer and viscosity respectively, $\sigma_{pa,sc} \approx 2 \cdot 10^{-18}$ m$^{2}$, $\sigma_{pa,mt}\approx 10^{-18}$ m$^{2}$, $\sigma_{pa,v}\approx 0.35 \cdot 10^{-18}$ m$^{2}$. For hydrogen the approximate values are $\sigma_{aa,sc}=\sigma_{aa,mt}\approx 10^{-18}$ m$^{2}$, and $\sigma_{aa,v}\approx 0.26\cdot 10^{-18}$ m$^{2}$.

    \begin{figure}%[!htb]
   \centering
  \includegraphics[height=6cm,bb=16 13 272 218,clip=]{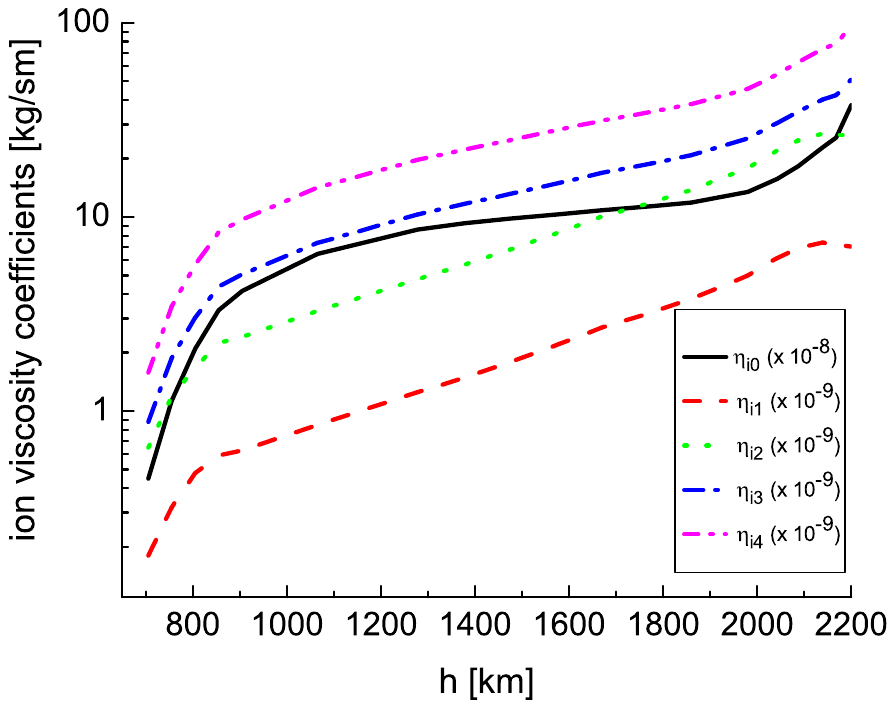}
      \caption{Proton  viscosity coefficients in chromosphere calculated from Eqs.~(\ref{vi}) and for altitude-dependent viscosity cross sections presented in Table~\ref{t1} (columns 5 and 7).   \label{fig1}}
       \end{figure}

Having the data for viscosity cross sections in Table~\ref{t1}, and using data from \citet{fon}, the ion viscosity coefficients are calculated from Eqs.~(\ref{vi}) and presented in Fig.~\ref{fig1}. Here we take a decreasing magnetic field model very similar to \citet{le}, in the shape
\be
B_0(h)=B_{00} \exp(-h/440), \label{b}
 \ee
 where $h$ is in km, and  $B_{00}=0.1$ T is the value at $h=0$.  So it changes from 0.016 T to 0.0007 T in the given altitude range in the figure.  Observe that $\eta_{i0}$ has values very similar to the curve $\eta_{pp}$ in Fig.~8 from \citet{vkr}; for example at $h=705$ km and $h=1065$ km presently we have $\eta_{i0}=0.45$ and $\eta_{i0}=6.46$, while $\eta_{pp}$ from \citet{vkr} has values 0.31, and 7.83, respectively, in the same units. These values are  remarkably close to each others, particularly  in view of the fact that $\eta_{pp}$ in the previous work  is obtained using the BGK collisional model integral for unmagnetized plasma. This  confirms that such a model integral is indeed able to yield results which coincide with expressions obtained from more advanced theory and  from exact calculations presented in \citet{zh}. From Fig.~\ref{fig1} it is seen that the coefficient $\eta_{i1}$ remains the smallest, while $\eta_{i4}$ becomes close to $\eta_{i0}$ above 2000 km where their ratio  is around 1/4. Generally, all coefficients are drastically reduced in photosphere, which is simply due to strong collisions. This   should be expected because $\eta_{i0}\sim 1/\nu_i$, i.e., viscosity is  suppressed by collisions, but this holds only up to some point because all derivations of viscosity theory are  based on the presence of collisions. These issues are nicely discussed by  \citet{cow}  and \citet{sh}.

    \begin{figure}%[!htb]
   \centering
  \includegraphics[height=6cm,bb=16 13 272 214,clip=]{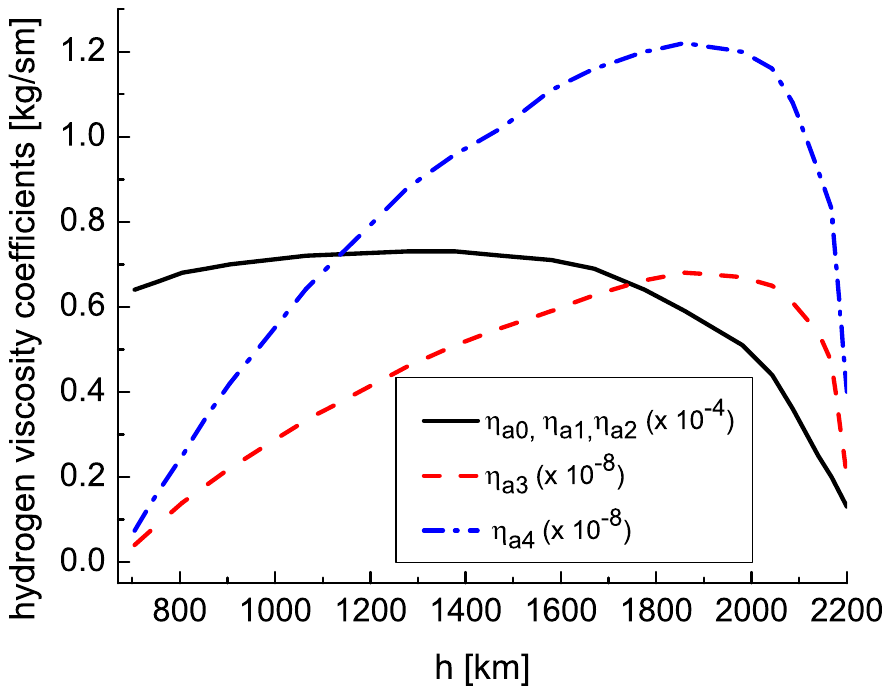}
      \caption{Atomic hydrogen viscosity coefficients in chromosphere calculated from Eqs.~(\ref{nv})  and for altitude-dependent viscosity cross sections presented in Table~\ref{t1} (columns 5 and 7).   \label{fig2}}
       \end{figure}

Viscosity coefficients for hydrogen atoms are presented in Fig.~\ref{fig2}, calculated from Eqs.~(\ref{nv}) and by using viscosity cross section given in Table~\ref{t1}. In the lower layers the coefficient $\eta_{a0}$ goes up to 0.7 in the given units, which is very similar to  $\eta_{{\sss HH}}$ given in \citet{vkr} in Fig.~11, where it is very slowly changing in the interval $\simeq 0.4-0.8$. But at higher altitudes  $\eta_{a0}$ decreases contrary to \cite{vkr}. One reason for the difference is clearly the following;  in \citet{vkr} in the expression for $\eta_{{\sss HH}}\sim 1/\sum_b \nu_{{\sss H} b}$, the summation is approximated by atom-atom collisions only, which is  valid in photosphere but may become inappropriate  in chromosphere in view of the altitude-dependent number densities where atom-ion collisions  after some $h$ become important or dominant, and this results in a  reduced friction coefficient with altitude. So this explains why in the present work $\eta_{a0}$ correctly decreases with $h$ while in \citet{vkr} it is slightly increased (due to reduced number of neutrals and less a-a collisions). Another possible reason for small differences  is most likely that  in the present work we adopted $A_{\alpha\beta}^*=1$, which is valid for  hard sphere model, while in \citet{vkr} a more advanced model for hydrogen cross section is used, so $A_{\alpha\beta}^*$ should be modified.
 But  calculating such tedious corrections for $A_{\alpha\beta}^*$ would be  redundant in view of  so insignificant differences.
All coefficients here are also decreased  at higher altitudes clearly because of smaller amount of neutrals which enter into $p_a$ in Eqs.~(\ref{nv}).

The viscosity coefficients presented in Figs.~\ref{fig1},~\ref{fig2} are for the given  magnetic field. Different field  values do not affect the leading order coefficients $\eta_{i0}, \eta_{a0, 1, 2}$, the others are affected. For example, taking $h=1065$ km and magnetic field 10 times stronger ($B_0=0.09$ T) yields $\eta_{i1,2}$ reduced by about two orders of magnitude and $\eta_{i3,4}, \eta_{a3,4} $ by one order of magnitude. On the other hand, reducing the magnetic field for one order of magnitude yields $b_i=8.65$ and all the values for  ion coefficients  become very close to the presently given $\eta_{i0}$, and this  may be expected from Eqs.~(\ref{vi}). So in strong magnetic structures the parallel and perpendicular viscosity effects may become drastically different. But actual role and effect of various components can be understood only bearing in mind the following: a) the viscosity coefficients are accompanied with second derivatives [see for example Eqs.~(\ref{d1}-\ref{d3}) related to the Alfv\'{e}n waves later in the text] that may have very  different values parallel and perpendicular to the magnetic field, and b) they are also accompanied with various (parallel and perpendicular) speed components that may be very different for different waves.

To get  some idea about parameters introduced in Sec.~\ref{s2a} we may take altitude $h=1065$ km, and for the corresponding  plasma parameters \citep{fon} we have
\[
\tau_i=10^{-5} \mbox{s}, \quad \tau_a=8\cdot 10^{-5} \mbox{s}, \quad \xi_i=1.45, \quad \xi_a=1,
\]
\[
  f_{ia}=0.1, \quad f_{ia}'=0.4, \quad a_{ia}=-0.0003, \quad a_{ai}=-0.45,
 \]
\[
  b_i=8.65,  \quad \Delta_{\eta}=0.99987, \quad L_{ep}=9.45,
\]
\[
 \tau_{ii}=3.9 \cdot 10^{-6} \mbox{s},  \quad \tau_{ei}=1.3 \cdot 10^{-7} \mbox{s}.
\]
The value for $b_i$ suggests that Alfv\'{e}n waves can be expected at this altitude, but  as   damped waves, and this will be checked in Sec.~\ref{s4}, see Figs.~\ref{fig3},\ref{fig4}.

\subsection{Viscosity coefficients for electrons in partially ionized plasmas }\label{s2b}

The electron viscosity is usually negligible in a three-component mixture with $T_i=T_a=T$, but this is both mass and temperature dependent. Since the mass ratio $\delta$ is usually known, it turns out that electron viscosity must be taken into account if  $T_e/T \simeq 4.5$ or greater \citep{zh}, because it becomes of the same order as the ion viscosity. The electron viscosity coefficients in a multicomponent partially ionized plasmas in magnetic field depend on the {\em ion magnetization} as well. The general expressions are  lengthy and they will not be given here as we are dealing with  the solar plasma where the temperature ratio is close or equal to unity so  that electron viscosity is negligible at least in application to the Alfv\`{e}n wave. But rather simple expressions may be given  in the limit $b_i\equiv \Omega_i \tau_i\ll 1$ (that is for unmagnetized ions), when electron viscosity decouples from viscosity of ions. Such a wide layer of unmagnetized ions and magnetized electrons in the photosphere and lower chromosphere is clearly identified by \cite{vkr}. So for such an environment, with the accuracy $\delta^{1/2} (T/T_e)^{5/2}$ the electron viscosity coefficients read \citep{zh}:
\[
\eta_{e0}=\frac{1}{2}n_e \kappa T_e \tau_e, \!\quad \eta_{e1}=\frac{\eta_{e0}}{1+ b_e^2}, \!\quad \eta_{e2}=\eta_{e1}\!\left[\frac{b_e}{2}\right]
\!\equiv\! \frac{\eta_{e0}}{1+ \frac{\displaystyle{b_e^2}}{\displaystyle{4}}},
\]
\[
 \eta_{e3}=b_e \eta_{e1}, \quad \eta_{e4}=\eta_{e3}\left[\frac{b_e}{2}\right]\equiv\frac{b_e}{2} \frac{\eta_{e0}}{1+ \frac{\displaystyle{b_e^2}}{\displaystyle{4}}}, \quad b_e=\Omega_e \tau_e,
\]
\[
\tau_e^{-1}=0.3 \tau_{ee}^{-1} + 0.6 \sum_{\beta\neq e}A_{e \beta}^* \tau_{e \beta}^{-1}.
\]
Summation in the last expression is over all species excluding  electrons and $A_{e\beta}^* \simeq 1$ for Coulomb collisions, with accuracy $1/L_{e\beta}$, and
$A_{e\beta}^* = 1$ for electron-atom collisions within the hard sphere atom model. Here we use  Eq.~(\ref{e3}) and $\tau_{ee}=\tau_{ei}/\sqrt{2}$, and $\tau_{ea}$ is described above in Sec.~\ref{s2a}.

In the other limit $b_i\geq 1$ numerous corrections appear in the coefficients  $\eta_{ej}$ \citep{zh}; for fully ionized plasma the expressions are less complicated and they are given in Sec.~\ref{s3cc}.

\section{Fully ionized electron-proton plasma }\label{s3}

Viscosity coefficients for fully ionized plasma are given here  for completeness and to show differences between Braginskii's results and more general results of Zhdanov. The latter are   completely overlooked by solar plasma researchers. We use  them later in the text to estimate the role of viscosity in the corona.
\subsection{Ion viscosity coefficients }\label{s3a}
\subsubsection{Results of Braginskii}
For arbitrary ratio $\Omega_i/\nu_i$ the viscosity coefficients for ions are  \citep{bra2}:
\be
\eta_{i0}=0.96 n_i \kappa T_i \tau_i, \quad \eta_{i2}= n_i \kappa T_i \tau_i \frac{1.2 b_i^2 + 2.33}{b_i^4 + 4.03 b_i^2 + 2.33}, \label{b1}
\ee
\be
 \eta_{i4}= n_i \kappa T_i \tau_i b_i \frac{b_i^2 + 2.38}{b_i^4 + 4.03 b_i^2 + 2.33},\quad b_i=\Omega_i \tau_i, \label{b2}
 \ee
  \[
\eta_{i1}= \eta_{i2}[2 b_i]= n_i \kappa T_i \tau_i \frac{1.2 (2 b_i)^2 + 2.33}{(2b_i)^4 + 4.03 (2b_i)^2 + 2.33},
\]
\[
\eta_{i3}=\eta_{i4}[2b_i]=n_i \kappa T_i \tau_i (2 b_i) \frac{(2b_i)^2 + 2.38}{(2b_i)^4 + 4.03 (2b_i)^2 + 2.33},
\]
Here, in view of comments given in Sec.~\ref{s2a} and using Eq.~(\ref{e2}) we have
\be
\tau_i\equiv \tau_b=\frac{12 \varepsilon_0^2 m_i^{1/2} (\pi \kappa T_i)^{3/2}}{e^4 z_i^4 n_i L_{ii}}= 2 \tau_{ii}.\label{e5}
\ee
\subsubsection{Results of Zhdanov}
Following \citet{zh}, the ion viscosity coefficients in case of electron and single ion plasma read:
\be
\eta_{i0}=1.92 n_i \kappa T_i \tau_{ii}, \quad \eta_{i2}= n_i \kappa T_i \tau_{ii}\frac{0.6 b_i^2 + 0.28}{b_i^4 + b_i^2 + 0.146}, \label{z1}
\ee
\be
 \eta_{i4}= n_i \kappa T_i \tau_{ii} b_i \frac{b_i^2 + 0.6}{b_i^4 + b_i^2 + 0.146}, \quad b_i=\Omega_i \tau_{ii}, \label{z2}
 \ee
\[
\eta_{i1}= \eta_{i2}[2 b_i], \quad  \eta_{i3}=\eta_{i4}[2b_i].
\]
Here, $\tau_{ii}$ is given by Eq.~(\ref{e2}). Bearing in mind the factor 2 difference in the collision time $\tau_i=2 \tau_{ii}$, where $\tau_i$ is the Braginskii's time (\ref{e5}) and $\tau_{ii}$ Zhdanov's time (\ref{e2}),  it is easy to see that the differences between the two sets of coefficients are negligible. Indeed, introducing subscript $zh$ for Zhdanov's parameters and subscript $br$ for those of Braginskii, and expressing $\tau_{ii}$ through $\tau_i$ in Zhdanov's results, we have $\eta_{i0, zh}=\eta_{i0, br}$.  Further noticing that for the same reasons  (i.e., the difference in definition of the ion collision time) $b_{i,zh}=b_{i, br}/2$, we can write Zhdanov's coefficient $\eta_{i2, zh}$ in terms of Braginskii's parameters as
\be
\eta_{i2, zh}=n_i \kappa T_i \tau_i \frac{1.2 b_{i,br}^2 + 2.24}{ b_{i,br}^4 + 4  b_{i,br}^2 + 2.336} \label{z3}.
\ee
Similarly for $\eta_{i4, zh}$ we have:
\be
\eta_{i4, zh}=n_i \kappa T_i \tau_i b_{i, br}\frac{b_{i,br}^2 + 2.4}{ b_{i,br}^4 + 4  b_{i,br}^2 + 2.336}. \label{z4}
\ee
Thus the comparison of these two sets of expressions  shows a remarkable similarity and insignificant differences (if the collision time is re-defined correspondingly) although they  are obtained following completely different procedures (though both based on some initial expansions); the Braginskii's Eqs.~(\ref{b1}, \ref{b2}) follow from the Chapman-Enskog scheme, while Zhdanov's Eqs.~(\ref{z1}, \ref{z2}) [or  Eqs.~(\ref{z3}, \ref{z4})]  are based on  the Grad's method.

\subsection{Electron  viscosity coefficients}\label{s3b}

Electron viscosity  plays no  role in application to AW in the solar plasma. But it may be of importance for some other modes even in solar plasma with $T_e=T_i$, like fast electron dynamics on the background of nearly static ions  in so called electron-MHD theory, or for electron waves in general within multicomponent theory. In the usual MHD theory the total viscosity is a sum of partial viscosities and electron contribution is thus  negligible.

\subsubsection{Braginskii's results}\label{s3bb}
For electron-proton plasma in magnetic field the viscosity coefficients read \citep{bra2}:
\be
\eta_{e0}=0.733 n_e \kappa T_e \tau_e, \quad \eta_{e2}\!=\!n_e \kappa T_e \tau_e \frac{2.05 b_e^2 + 8.5}{b_e^4\! +\! 13.8 b_e^2\! +\! 11.6}, \label{b22}
\ee
\be
\eta_{e4}=- n_e \kappa T_e \tau_e b_e \frac{b_e^2 + 7.91}{b_e^4 + 13.8 b_e^2 + 11.6}, \quad \eta_{e1}=\eta_{e2}[2 b_e], \label{b23}
\ee
\[
\eta_{e3}=\eta_{e4}[2 b_e],\quad b_e=\Omega_e\tau_e.
\]
Here, in Braginskii's notation $\tau_e$ coincides (after rewriting his expression in SI units) with the earlier given $\tau_{ei}$ in Eq.~(\ref{e3}), and  Braginskii uses $L_e$ (instead of $L_{ei}$)  given as $L_e=23.4 - 1.15 \log[n_e]+ 3.45 \log[T_e]$ for $T_e< 50 $ eV, and $L_e=25.3 -1.15 \log[n_e] + 2.3 \log[T_e]$ for $T_e>50$ eV.

\subsubsection{Zhdanov's results}\label{s3cc}
For general plasma with ions having the charge number $z_i$ the viscosity coefficients are:
\[
\eta_{e0}=n_e \kappa T_e \tau_{ei} \frac{s_1}{s_0}, \quad \eta_{e2}=n_e \kappa T_e \tau_{ei} \frac{s_1 s_0+ s_2 b_e^2}{s_0^2 + s_4 b_e^2 + b_e^4},
\]
\[
\eta_{e4}=-n_e \kappa T_e \tau_{ei} b_e \frac{s_3 + b_e^2}{s_0^2 + s_4 b_e^2 + b_e^4}, \quad \eta_{e1}=\eta_{e2}[2 b_e],
\]
\[
\eta_{e3}=\eta_{e4}[2 b_e],\quad b_e=\Omega_e\tau_{ei}, \quad s_0=0.82 + \frac{1.82}{z_i}+ \frac{0.72}{z_i^2},
\]
\[
 s_1=1.46 + \frac{1.04}{z_i},\quad s_2=1.2 + \frac{0.85}{z_i},
 \]
\[
s_3=3.05 + \frac{3.7}{z_i} + \frac{1.17}{z_i^2}, \quad s_4=5.32 + \frac{6.36}{z_i} + \frac{2.02}{z_i^2}.
\]
For electron-proton plasma with $z_i=1$ this yields
\be
\eta_{e0}\!=\!0.744 n_e \kappa T_e \tau_{ei}, \! \quad \eta_{e2}\!=\!n_e \kappa T_e \tau_{ei} \frac{2.05 b_e^2 + 8.4}{b_e^4\! +\! 13.7 b_e^2 \!+\! 11.29}, \label{z22}
\ee
\be
\eta_{e4}\!=\!-n_e \kappa T_e \tau_{ei} b_e \frac{ b_e^2 + 7.92}{b_e^4 + 13.7 b_e^2 + 11.29}. \label{z23}
\ee
Since $b_e$ here and in Braginskii's expressions in Sec.~\ref{s3bb} are the same, we see again that  Eqs.~(\ref{b22}, \ref{b23}) from one side, and Eqs.~(\ref{z22}, \ref{z23}) from the other,  demonstrate an extraordinary agreement in results between the two completely different methods. %In the same time it %should be pointed out again that  the results from \citet{zh} are much more general.

\section{Application to  waves}\label{s4}
We shall use the following set of linearized momentum equations for electrons,  ions,  and neutral atoms, which can be applied  for various plasma waves:
\[
m_en_0 \frac{\partial \vec v_e}{\partial t}  =-\nabla p_e - \nabla\!\cdot\! \Pi_e   -e n_0\left(\vec E + \vec v_e\times \vec B_0\right)
\]
\be
- m_e n_0 \nu_{en} (\vec v_e - \vec v_n) - m_e n_0 \nu_{ei} (\vec v_e - \vec v_i),\label{si2}
 \ee
\[
m_in_0\frac{\partial \vec v_i}{\partial t}\!  = -\nabla p_i - \nabla\!\cdot\! \Pi_i + e n_0\left(\vec E + \vec v_i\times \vec B_0\right)
\]
\be
+ m_e n_0 \nu_{ei} (\vec v_e - \vec v_i) - m_i n_0 \nu_{in} (\vec v_i - \vec v_n),\label{si1}
 \ee
 \[
m_n n_n \frac{\partial \vec v_n}{\partial t}  =  -\nabla p_n - \nabla\!\cdot\! \Pi_n   + m_e n_0 \nu_{en} (\vec v_e - \vec v_n)
\]
\be
+ m_i n_0 \nu_{in} (\vec v_i - \vec v_n).
 \label{si3}
\ee
Not all terms given here are of equal importance for every wave and this will be demonstrated below in application to  Alfv\'{e}n  waves (AW) and electron plasma (EP) waves.

\subsection{Alfv\'{e}n  waves}\label{s41}

\subsubsection{Chromosphere}\label{avcr}

In case of perturbations  propagating along the magnetic field $\vec k=k\vec e_x$ (dependence on the $x$-coordinate only), the components $W_{\alpha rs}$ become:
\[
W_{\alpha xx} =\frac{4}{3} \frac{\partial v_{\alpha x}}{\partial x},   \quad  W_{\alpha xy} = W_{\alpha yx}= \frac{\partial v_{\alpha y}}{\partial x},
 \]
 \[
  W_{\alpha xz} = W_{\alpha zx}= \frac{\partial v_{\alpha z}}{\partial x},\quad W_{\alpha yy} =  W_{\alpha zz} = -\frac{2}{3} \frac{\partial v_{\alpha x}}{\partial x},
 \]
 \be
 W_{\alpha yz}= W_{\alpha zy}=0. \label{w1}
\ee
The  viscosity tensor for ions and neutrals becomes:
\begin{equation}
\Pi_{\alpha}\!=\!\left(\!\!
  \begin{array}{ccc}
    \vspace{0.15cm}
    -\frac{\displaystyle{4 \eta_{\alpha 0}}}{\displaystyle{3}} \frac{\displaystyle{\partial v_{\alpha x}}}{\displaystyle{\partial x}} &
     \Pi_{xy} &  \Pi_{xz} \\
     \vspace{0.15cm}
   \Pi_{yx} &  \frac{\displaystyle{2 \eta_{\alpha 0}}}{\displaystyle{3}} \frac{\displaystyle{\partial v_{\alpha x}}}{\displaystyle{\partial x}} & 0 \\
       \Pi_{zx} & 0 &   \frac{\displaystyle{2 \eta_{\alpha 0}}}{\displaystyle{3}} \frac{\displaystyle{\partial v_{\alpha x}}}{\displaystyle{\partial x}} \\
   \end{array}
\!\!\right), \label{pi1a}
\end{equation}
\[
 \Pi_{xy}= \Pi_{yx}=-\eta_{\alpha 2} \frac{\displaystyle{\partial v_{\alpha y}}}{\displaystyle{\partial x}}-  \eta_{\alpha 4} \frac{\displaystyle{\partial v_{\alpha z}}}{\displaystyle{\partial x}},
 \]
\[
 \Pi_{xz}= \Pi_{zx}= -\eta_{\alpha 2} \frac{\displaystyle{\partial v_{\alpha z}}}{\displaystyle{\partial x}}+   \eta_{\alpha 4} \frac{\displaystyle{\partial v_{\alpha y}}}{\displaystyle{\partial x}} \quad \alpha=i, a.
 \]
The contribution of the stress tensor in the parallel $x$ and   perpendicular $y, z$  components in momentum equation are:
\be
\left(\nabla \cdot \Pi_{\alpha}\right)_x=-\frac{4 \eta_{\alpha 0}}{3} \frac{\partial^2 v_{\alpha x}}{\partial x^2}, \label{d1}
\ee
\be
\left(\nabla \cdot \Pi_{\alpha}\right)_y=-\eta_{\alpha 2} \frac{\partial^2 v_{\alpha y}}{\partial x^2} - \eta_{\alpha 4} \frac{\partial^2 v_{\alpha z}}{\partial x^2}, \label{d2}
\ee
\be
\left(\nabla \cdot \Pi_{\alpha}\right)_z=-\eta_{\alpha 2} \frac{\partial^2 v_{\alpha z}}{\partial x^2} + \eta_{\alpha 4} \frac{\partial^2 v_{\alpha y}}{\partial x^2}. \label{d3}
\ee
 Without magnetic field the tensor components $\Pi_{ars}$ reduce to  $\Pi_{ajj}$, $j\in x, y, z$, with the only remaining coefficient $\eta_{a0}$ given earlier in Eq.~(\ref{nv}). However,  magnetic field affects the viscosity tensor of neutrals, which now  contains  7 non-zero components just like the ion tensor.

Combined  Faraday law $\nabla\times \vec E_1=- \partial \vec B_1/\partial t$ and  Amp\`{e}re law $\nabla \times \vec B= \mu_0 \vec j$ without displacement current yield the general wave equation:
\be
\nabla \times \left(\nabla \times \vec E_1\right) + \mu_0 \frac{\partial \vec j_1}{\partial t}=0. \label{aw1}
\ee
Within the linear theory  $\vec j_1= e n_0 (\vec v_{i1}-\vec v_{e1})$,  and for the background magnetic field $\vec B_0=B_0\vec e_x$ and transverse electromagnetic perturbations propagating along the background field $\sim\exp(-i \omega t+ i k x)$, the wave equation becomes
\be
k^2  \vec E_1-i \mu_0\omega e n_0(\vec v_{i1}-\vec v_{e1})=0. \label{aw2a}
\ee
  The electric field may be assumed with the  component in $y$-direction, and in this case  the dispersion equation for the Alfv\'{e}n wave reads \citep{vko}
\be
\omega^2=\frac{k^2 c_a^2}{1+ k^2\lambda_i^2}, \quad \lambda_i=\frac{c}{\omega_{pi}}. \label{f1}
\ee
The term in denominator is small but  it can be shown that it comes from the ion speed in the $y$-direction (the polarization drift), and if the polarization drift was omitted in the first place there would be no AW at all. So the term has a clear  physical meaning and should be kept in derivations.
 Observe also that $k\lambda_i=k c_a/\Omega_i\simeq \omega/\Omega_i $, so small  $k\lambda_i$  implies the  well-known AW frequency limit $\omega/\Omega_i\ll 1$. As an example, in chromosphere at $h=1065$ km, $\lambda_i=0.7$ m; at $h=1980$ km it is $1.1$ m, so discussing AW in this environment implies scales much exceeding these lengths.

In the assumed geometry $\vec B_1=B_1\vec e_z$, from the Faraday law   we have $\vec E_1=E_1\vec e_y$, and this yields
\be
k^2  E_x-i \mu_0\omega e n_0(v_{iy}-v_{ey})=0. \label{aw2}
\ee
Here  $v_{iz}=v_{ez}$ which is just the leading order $\vec E_1\times \vec B_0$-drift equal for both species. From Eqs.~(\ref{si2}, \ref{si1}) it is clear that within the linear theory the parallel dynamics
is decoupled, so all we need are $y, z$ speed components from the tree momentum equations. We shall thus keep  viscosity effects for ions and neutrals through Eqs.~(\ref{d2}, \ref{d3}), and the pressure terms of all species are omitted for the simple shear  Alfv\'{e}n wave dynamics.

\begin{figure}%[!htb]
   \centering
  \includegraphics[height=6cm,bb=16 14 281 218,clip=]{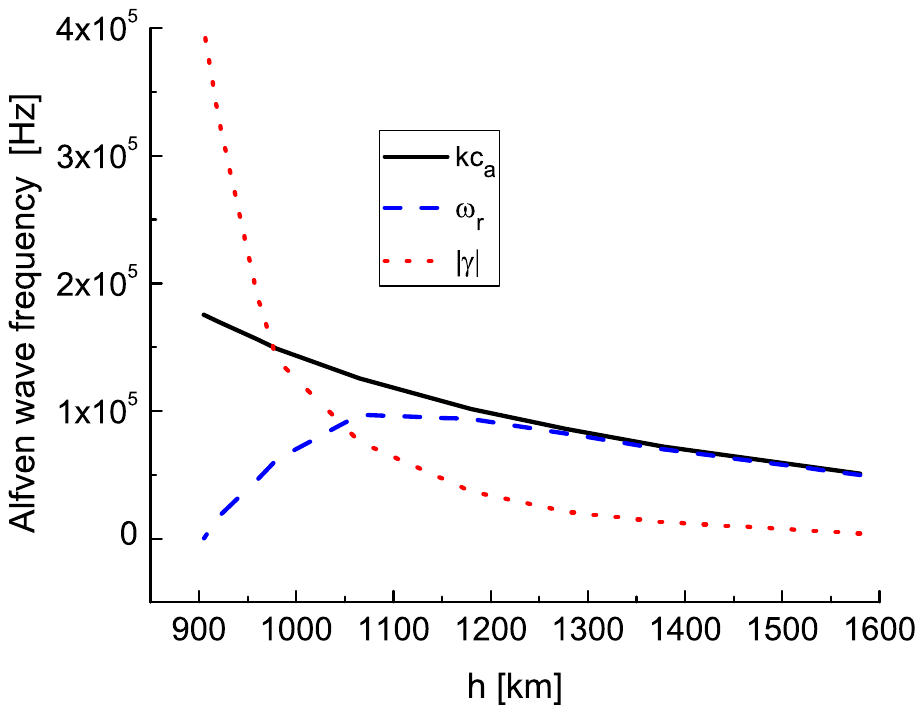}
      \caption{Alfv\'{e}n wave frequency with altitude for wavelength $\lambda=30$ m.   \label{fig3}}
       \end{figure}

Eqs.~ (\ref{si2}-\ref{si3}, \ref{aw2}) yield the following  set of seven equations for $v_{ey}, v_{ez}, v_{iy}, v_{iz}, v_{ny}, v_{nz}, E_1$:
\be
\left(\!\!
        \begin{array}{ccccccc}
           a_{11} &  a_{12} &  a_{13} &  0 &  a_{15} &  0 &  e \\
           a_{21} &  a_{22} &  0 &  a_{24} &  0 &  a_{26} &  0 \\
           a_{31} &  0 &  a_{33} &  a_{34} &  a_{35} &  0 &  e \\
          0 &  a_{42} &  a_{43} &  a_{44} &  0 &  a_{46} &  0 \\
           a_{51} &  0 &  a_{53} &  0 &  a_{55} &  a_{56} &  0 \\
           0 &  a_{62} &  0 &  a_{64} &  a_{65} &  a_{66} &  0 \\
           a_{71} &  0 &  a_{73} &  0 &  0 &  0 &  -k^2 \\
        \end{array}
    \! \! \right)\! \!\! \left(\!\!
                 \begin{array}{c}
                   v_{ey} \\
                   v_{ez} \\
                   v_{iy} \\
                   v_{iz} \\
                   v_{ay}\\
                   v_{az} \\
                   E_1 \\
                 \end{array}\!\!\right)\!=\!0, \label{aw3}
\ee
\[
 a_{11}=m_e(-i\omega + \nu_{ei} + \nu_{en}), \quad  a_{12}=eB_0, \quad  a_{13}=-m_e \nu_{ei},
 \]
\[
 a_{15}=-m_e\nu_{en}, \quad  a_{21}\!=eB_0, \!\quad a_{22}\!= m_e(i\omega - \nu_{ei} - \nu_{en}),
 \]
\[
a_{24}=m_e \nu_{ei}, \quad a_{26}=m_e \nu_{en}, \quad a_{31}=m_e \nu_{ei},
\]
\[
 a_{33}= i m_i \omega - m_e \nu_{ei}- m_i \nu_{in} -\frac{\eta_{i2}k^2}{n_0},\!\quad a_{34}= e B_0-\frac{\eta_{i4}k^2}{n_0},
\]
\[
a_{35}=m_i \nu_{in}, \quad   a_{42}=m_e \nu_{ei}, \quad a_{43}=- a_{34}, a_{44}= a_{33},
\]
\[
a_{46}=m_i\nu_{in},  \quad a_{51}=m_en_0 \nu_{en},  \quad a_{53}=m_i n_0 \nu_{in},
\]
\[
a_{55}= i \omega m_n n_n - m_e n_0 \nu_{en} - m_i n_0 \nu_{in}-\eta_{a2} k^2\!,
\]
\[
a_{56}=-\eta_{a4} k^2\!, \quad a_{62}=m_e n_0 \nu_{en}, \quad  a_{64}=m_i n_0 \nu_{in},
\]
\[
 a_{65}= \eta_{a4} k^2, \quad  a_{66}= a_{55}, \quad a_{71}=  -i \mu_0 \omega e n_0=-a_{73}.
\]
Dispersion equation is obtained by setting the $[7\times 7]$ matrix on the left-hand side in (\ref{aw3}) equal to zero,
\be
\Delta_{7\times 7}(\omega, k)=0. \label{aw4}
\ee
This dispersion equation is solved numerically and as one example  Fig.~\ref{fig3} shows  the  calculated wave frequency $\omega_r$  and the corresponding damping rate $-\gamma$ for a chosen   wavelength $\lambda=30 $ m. The full line in the figure depicts the value $k c_a$, given here just for comparison with an idealized situation without collisions. The increase  at  low  altitudes is due to the increased magnetic field as described by equation (\ref{b}). But in the given realistic situation, it is seen that below 1000 km the wave is heavily damped and its frequency $\omega_r$ vanishes completely before reaching the altitude of 900 km and reduces to zero. Observe that at $h=905$ km, the ion inertial length $\lambda_i=0.7$ m, and $k \lambda_i=0.14$ so clearly going to shorter wavelengths violates the assumptions incorporated into the theory. At $h=1580$ km the wave is very weakly damped, $|\gamma|/\omega_r=0.08$, and above this layer the ratio is even lower. Thus, in these layers the dispersion line is close to the ideal case and the wave is expected to exist and to propagate.

  Both cases with and without viscosity are checked and differences in the presented graph are invisible, the viscosity plays no role at all. This may be understood  also by directly comparing the leading remaining viscosity term $\eta_{i4} k^2v_{i1}=R_{vi} v_{i1}$ and the ion-neutral friction term $m_i n_0\nu_{in} (v_{i1}-v_{n1})$. In case of static neutrals (e.g., for relatively short wavelengths when they merely represent an obstacle, or at an initial stage of the perturbation when charged particles are set into motion first), the latter  may be written as $\simeq R_{fi} v_{i1}$. At $h=905$ km the ratio is $R_{vi}/R_{fi}\simeq 4\cdot 10^{-6}$; at $h=1860$ km it is  $R_{vi}/R_{fi}\simeq 0.01$, so for the ion dynamics the viscosity is negligible in all the altitude range. This is expected to remain so even for larger wavelengths in view of the $k^2\!$-dependence of the viscosity. This issue is discussed in detail in \citet{vkr}.

\begin{figure}%[!htb]
   \centering
  \includegraphics[height=6cm,bb=16 13 270 218,clip=]{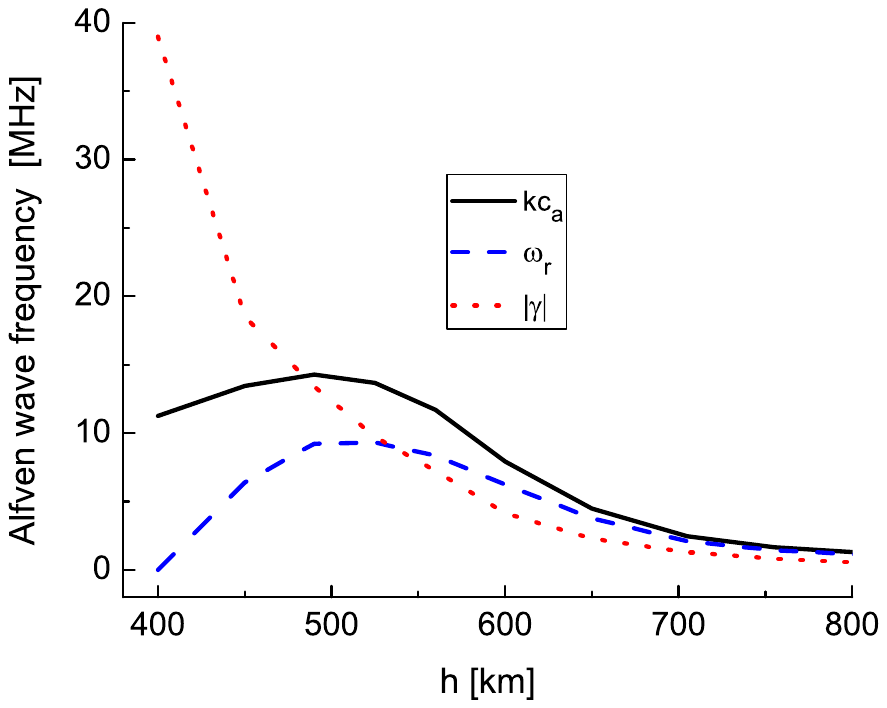}
      \caption{Alfv\'{e}n wave frequency with altitude for wavelength $\lambda=30$ m and magnetic field 5 times stronger as compared to Fig.~\ref{fig3}.   \label{fig4}}
       \end{figure}

The mode is sensitive to the value of $B_{00}$ which is used as given by Eq.~(\ref{b}). Taking the field five  times stronger  moves the layer of evanescence down to close to $h=400$ km when the mode frequency is zero, as presented in Fig.~\ref{fig4} for the same wavelength $\lambda=30$ m. The damping is huge in all the altitude range and although the mode appears formally possible in the given range and for such a strong field, in practical terms it will never appear. The magnetic field in the figure is 0.2 T at 400 km and decreases as described by Eq.~(\ref{b}).

Similarly, taking the magnetic field five times weaker the mode becomes evanescent in all the layers up to around $h=1200$ km. These all results are completely in agrement with \citet{vko}.
 \begin{figure}%[!htb]
   \centering
  \includegraphics[height=6cm,bb=16 13 270 218,clip=]{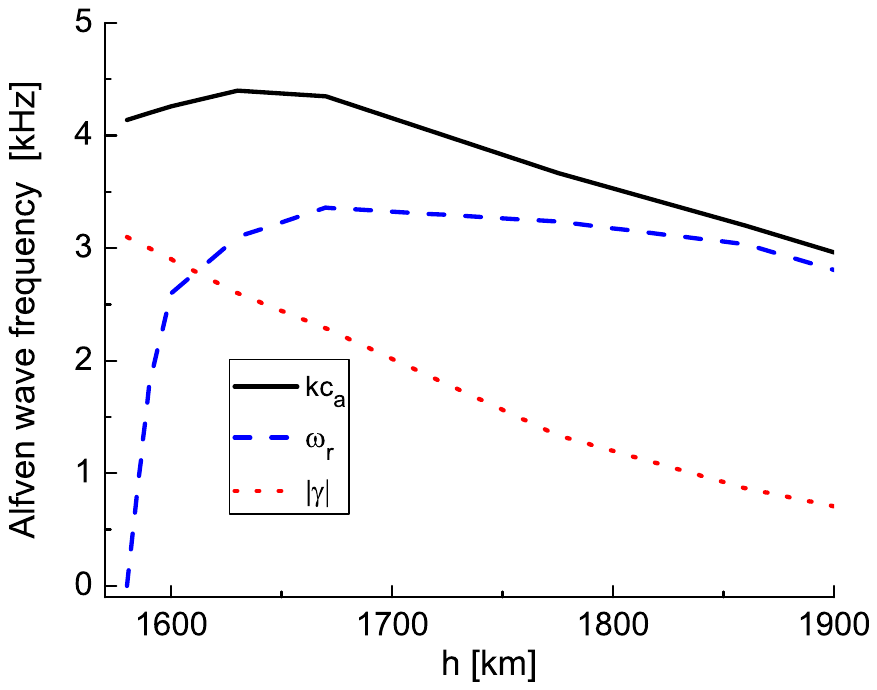}
      \caption{Alfv\'{e}n wave frequency with altitude for wavelength $\lambda=300$ m and magnetic field same as in  Fig.~\ref{fig3}.   \label{fig5}}
       \end{figure}
For longer wavelengths the situation regarding AW excitation is much worse because of an increased number of collisions within the wave period as discussed by   \citet{vko} for photosphere. Very similar is wave behavior  for the magnetized plasma in the present work. This is seen from Fig.~\ref{fig5} for $\lambda=300$ m where the magnetic field is the same as in Fig.~\ref{fig3}, given by Eq.~(\ref{b}). The mode is strongly damped below 1600 km and it completely vanishes at around 1580 km. As in the previous two figures the features of the presented ideal mode $k c_a$ are completely different.

Such relatively short wavelengths studied here are of particular importance because of an efficient coupling of the Alfv\'{e}n wave with the drift wave in an inhomogeneous environment, like  in  the magnetic loops in the solar atmosphere as shown by \citet{vapj}. In that study a strong  stochastic heating is obtained in magnetic loops, and it was shown to be much stronger  in regions with a stronger magnetic field. The energy for the wave growth and consequent heating is stored in the plasma inhomogeneity (i.e., in the gradient of density, temperature, and magnetic field). It is found that the energy release rate caused by the stochastic heating can be several orders of magnitude above
the value presently accepted as necessary for a sustainable heating. The vertical stratification and the very
long wavelengths along the magnetic loops imply that a drift-Alfv\'{e}n wave, propagating as a twisted structure along
the loop, in fact occupies regions with different plasma-beta and, therefore, may have different electrostatic-electromagnetic  properties,
resulting in different heating rates within just one or two wavelengths.

\subsubsection{Corona}\label{avco}
With the presented formulas we may now estimate the role of viscosity in corona even without solving dispersion equation for this environment.  Adopting $n_0=10^{15}$ m$^{-3}$ and $T=10^6$ K and $B_0=5\cdot 10^{-4}$ T, and assuming absence of neutrals, the only ion friction is with electrons. The relevant collision frequencies are $\nu_{ie}=0.038$ Hz, and $\nu_{ii}= 2.3$ Hz.

 We use Zhdanov's expressions for fully ionized plasma given in Sec.~\ref{s3a}. According to Eqs.~(\ref{d2}, \ref{d3})  we need $\eta_{i2}, \eta_{i4}$ and their values  are $\eta_{i2}=8.4\cdot 10^{-12}$ kg/(sm), $\eta_{i4}=2.9\cdot 10^{-7}$ kg/(sm). Observe also that $\eta_{i0}=0.0114 $ kg/(sm) is much larger due to the fact that the other coefficients are strongly affected by  the magnetic field, because ions are strongly magnetized $b_i=20610$ for the given value of $B_0$ and other parameters. In the friction term we have $\vec v_i-\vec v_e$ which can only be polarization drift speeds because the other, leading order $\vec E\times\vec B$-drifts, are equal. The polarization drift is mass dependent \citep{vko}, $v_{pj}=(1/\Omega_jB_0)\partial E_1/\partial t$, therefore the electron part can be omitted. Using $E_1=\omega B_1/k$ this yields $v_{pi}=[\omega^2/(k \Omega_i)](B_1/B_0)$ and  $R_{fi}=m_i n_0 \nu_{ie} v_{pi}$.   Adopting one percent perturbed magnetic field yields the viscosity/friction ratio  $R_{vi}/R_{fi}>1 $ for wavelengths $\lambda<\lambda_{max}=1.1$ km.

Taking the field $B_0=10^{-4}$ T yields $R_{vi}/R_{fi}>1 $ for wavelengths $\lambda<\lambda_{max}=28$ km. These data reveal  scales above which keeping viscosity would be redundant.

This issue could be discussed from the electron momentum equation as well. Due to the reasons explained above we will have $F_{fe}=F_{fi}$ because $m_e \nu_{ei}=m_e \nu_{ie}$. For the electron viscosity we use the expressions of Zhdanov from Sec.~\ref{s3b}. For  $B_0=5\cdot 10^{-4}$ the coefficients we need are $ \eta_{e2}=2.7\cdot 10^{-16}$ kg/(s m), and $ \eta_{e4}=-1.6\cdot 10^{-10}$ kg/(s m), and $\nu_{ei}=70.4$ Hz.  Hence, $R_{ve}/R_{fe}>1$ for $\lambda<0.6$ m. It may be concluded that for the AW the electron viscosity plays no role in the corona.

\subsection{Electron plasma waves}\label{s42}

\subsubsection{Chromosphere}\label{s422}
A completely different  example is the EP wave, in particular for perturbations propagating along the magnetic field vector when the magnetic effects play no important role. For this mode it is appropriate to keep electron pressure and viscosity tensor effects, while on such fast scales the ions and neutrals may be assumed as a static background. Without analyzing waves in detail (no doubt it is also strongly damped) we can estimate the relative importance of dissipative effects, i.e., the electron viscosity in comparison with their friction with the two heavy species. For this purpose we may now use expressions given in Sec.~\ref{s2b}, and in particular the leading order term $\eta_{e0}$.
\begin{table}
% \centering
% \begin{minipage}{140mm}
  \caption{The maximal wavelength of the electron plasma wave for which the electron viscosity dominates combined dissipations caused by friction with ions and neutrals.}
  \label{t2}
\begin{tabular}{|l|l|l|l|l|l}
  \hline
 $ h$ [km] &  $400$  & $755$ & $1180$ & $1860$ & $2200$ \\
   % \hline
 $\lambda_{max} $ [m] &$0.0025$  & $0.06$ & $0.2$ & $0.6$ & $2.3$  \\
   \hline
\end{tabular}
%\end{minipage}
\end{table}
The viscosity term in Eq.~(\ref{si2})  yields the following leading order term $4 \eta_{e0} k^2 v_{ex1}/3=R_{ve} v_{ex1}$ [see Eq.~(\ref{d1})]. The two friction terms are  $m_e n_0\nu_{en} v_{ex1}=R_{fe1} v_{ex1}$ and $m_e n_0\nu_{ei} v_{ex1}=R_{fe2} v_{ex1}$. The Fig.~5  for electron collision frequency in \citet{vkr} shows that the role of e-a and e-i collisions is altitude dependent, and in the same time the relative importance  of these two friction effects changes drastically. In fact, electron-atom collisions are bay far more dominant up to the altitude $h\simeq 900$ km, and above this layer e-i collisions dominate. Their total contribution we shall express through $R_{fe}=R_{fe1}+ R_{fe2}$.

Hence, we make the ratio $R_{ve}/R_{fe}$ and check its values with altitude for various wavelengths using Table~\ref{t1} and the formulas given before. In Table~\ref{t2} we give several  altitudes and maximal wavelengths $\lambda_m$ for which the viscosity is more dominant than the combined friction. For larger wavelengths the electron viscosity is negligible.

The wave may be damped due to purely kinetic effects as well. However, in case of a strongly collisional plasma like in the lower solar atmosphere the  particle-wave resonance is inefficient [one example of this effect for the acoustic mode has been studied by \citet{ok}], and kinetic effects are not expected to play an important role, but this is not so for higher layers.

\subsubsection{Corona}\label{s423}
Adopting $n_0=10^{15}$ m$^{-3}$ and $T=10^6$ K, for corona we obtain $\nu_{ei}=70.4$ Hz. Using Zhdanov's expressions for fully ionized plasma from Sec.~\ref{s3cc} we have $\eta_{e0}=1.46\cdot 10^{-4}$ kg/(sm).
The ratio $R_{ve}/R_{fe}$ now yields that the viscosity is dominant up to wavelengths $\lambda_{max}=340$ km, which in practice means always for this kind of waves with short wavelengths.

In  such an environment  the collision frequency is reduced and it may be appropriate to check the kinetic effects, like the Landau damping. For the wave frequency $\omega\approx \omega_{pe}=1.8\cdot 10^9$ Hz the normalized Landau damping (without collisions) becomes $\gamma_{nor}=\gamma/\omega_{pe}\approx -(\pi/8)^{1/2} f(y)$, where $y=k \lambda_{de}$, $\lambda_{de}=\vte/\omega_{pe}$, and $f(y)=1/\{y^3 \exp[1/(2 y^2)+ 3/2]\}$. This expression is valid for $k \lambda_{de}\ll 1$ and for $\omega\gg k \vte$. The shape of the function $f(y)$ is such that it has a strong extremum  localized at around $y\approx 0.6$.
The Debye radius $\lambda_{de}$ in the given case  is around $10^{-3}$ m, and this maximum damping is at wavelengths of around 0.15 m; for much different values it is negligible. But the conditions used above to get this analytical expression for damping  are such that we are far from the extremum and the damping for such wavelengths is completely negligible. So to calculate actual Landau damping for any wavelength it would be necessary to solve numerically the dispersion equation which contains the integral plasma dispersion function. However, this all is the matter of the kinetic theory and it is far beyond the scope of the present work which is aimed only at establishing the leading dissipative mechanism within the fluid theory. In the past this  fluid theory has been widely used in application to the coronal plasma, and the two dissipative effects have been  used in a rather arbitrary manner. We have shown that within the same fluid theory it is only viscosity that matters.

\section{Summary and conclusions}\label{s5}
In this work complete and self-consistent viscosity coefficients are presented for partially and fully ionized plasma and applied to the lower solar atmosphere and corona in analysis of Alfv\'{e}n and Langmuir waves as most different modes regarding spatial and temporal scales. It is shown that in photosphere and chromosphere viscosity has no practical importance for the Alfv\'{e}n  wave which is however heavily affected by friction.  The results presented in Sec.~\ref{s4} are in general agreement with our previous results \citet{v07,v08,vko}. In the upper chromosphere Alfv\'{e}n waves may propagate although typically as very damped modes. So some efficient drivers are instabilities are required to have these  and other waves present there in order to take part in the heating  which still remains a puzzle  \citep{vm1, vm2, vm3, vm4, vaa11, pw, pw13}. In the corona the Alfv\'{e}n  wave is affected by viscosity for wavelengths roughly below the limit of around 30 km.

For the Langmuir wave the interplay between viscosity and friction in photosphere and chromosphere is summarized in Table~\ref{t2}. In the corona the viscosity is practically always dominant.

The result presented for partially ionized plasma are based on the most accurate cross sections which include several essential effects simultaneously, i.e., quantum mechanical indistinguishability at low energies, charge exchange and atom polarization in the field of an external charge. All relevant cross sections are presented in Table~\ref{t1}. It shows that in spite of the temperature change in the range of about 4400 K to 13500 K in most situations one can use some mean values for the cross sections whose values are suggested in the text.

\appendix

\section[]{Collision cross sections with altitude (temperature) }\label{ap1}
Cross sections given in Table~\ref{t1} are obtained by using data from \citet{kr1, kr2} with recalculated energies (temperatures) from center of mass frame, as given in these references, to lab frame \citep{vkr}. The visible nearly periodic variations  are  a common feature at so low energies. Due to this,  the effects of the temperature minimum at $h=525$ km are not pronounced at all, see also the graphs  in \citet{vkr}. The calculations presented in \citet{kr1} and in \citet{kr2} are  based on quantum-mechanical principle of indistinguishability which must be taken into account at energies below 1 eV. This effect is included in all $pa$ cross sections. In the same time, charge exchange effect is included in $\sigma_{pa,sc}, \sigma_{pa,mt}, \sigma_{pa,v}$ \citep{vkr}.

 For atom-atom collisions  both direct and recoil scattering is accounted for, as a direct consequence of the indistinguishability
of particles \citep{vkr}. As  a result,  the presented values for $aa$ collisions are twice as high as the classic values obtained from the model of distinguishable particles.

For so narrow temperature (altitude) range, the normally present monotonously decreasing profile for cross sections is only seen for   $\sigma_{aa,v}$ but it is very weak, from $0.288$ at temperature minimum to $0.223$ at $h=2200$ km; all others show only the usual low-energy variations.
\begin{table*}
% \centering
% \begin{minipage}{140mm}
  \caption{Proton-hydrogen $pa$ and hydrogen-hydrogen $aa$ cross sections for scattering $sc$, momentum transfer $mt$, and viscosity $v$ in terms of  altitude (i.e., of temperature) in photosphere and chromosphere, all in units  $10^{-18}$ m$^2$.}
  \label{t1}
\begin{tabular}{|l|l|l|l|l|l|l|l}
  \hline
 $ h$ [km] & $ T $ [K] & $\sigma_{pa,sc}$  & $\sigma_{pa,mt}$ & $\sigma_{pa,v}$ & & $\sigma_{aa,sc, mt}$ & $\sigma_{aa,v}$ \\
    \hline
 0 & 6520 & 1.864 & 1.105 & 0.376   & &1.173 &0.271  \\
50 & 5790 & 2.102 & 1.109 & 0.357   & &1.138 &0.273 \\
100 & 5410 & 2.028 & 0.957 & 0.401  & &1.094 &0.276 \\
150 & 5150 & 2.173 & 0.973 & 0.400  & &1.068 &0.279 \\
200 & 4990 & 2.269 & 1.040 & 0.395  & &1.055 &0.281 \\
250 & 4880 & 2.307 & 1.080 & 0.392  & &1.048 &0.282 \\
300 & 4770 & 2.338 & 1.143 & 0.382  & &1.043 &0.283 \\
350 & 4660 & 2.201 & 1.047 & 0.388  & &1.041 &0.290 \\
400 & 4560 & 2.115 & 0.956 & 0.411  & &1.040 &0.286 \\
450 & 4460 & 2.125 & 0.997 & 0.425  & &1.042 &0.287 \\
490 & 4410 & 2.145 & 0.996 & 0.428  & &1.044 &0.288 \\
525 & 4400 & 2.159 & 1.031 & 0.426  & &1.045 &0.288 \\
560 & 4430 & 2.145 & 0.998 & 0.411  & &1.044 &0.287 \\
600 & 4550 & 2.115 & 0.956 & 0.412  & &1.041 &0.286 \\
650 & 4750 & 2.338 & 1.143 & 0.382  & &1.043 &0.283 \\
705 & 5030 & 2.269 & 1.040 & 0.395  & &1.058 &0.280 \\
755 & 5280 & 2.121 & 0.955 & 0.402  & &1.080 &0.277 \\
805 & 5490 & 2.008 & 0.998 & 0.392  & &1.103 &0.275 \\
855 & 5650 & 2.092 & 1.144 & 0.367  & &1.122 &0.274 \\
905 & 5755 & 2.102 & 1.154 & 0.357  & &1.134 &0.273 \\
980 & 5900 & 2.004 & 1.060 & 0.368  & &1.145 &0.272 \\
1065& 6040 & 1.948 & 1.025 & 0.373  & &1.151 &0.272 \\
1180& 6230 & 1.864 & 1.006 & 0.379  & &1.157 &0.271 \\
1278& 6390 & 1.841 & 1.054 & 0.381  & &1.160 &0.270 \\
1378& 6560 & 1.924 & 1.173 & 0.365  & &1.161 &0.269 \\
1475& 6720 & 2.000 & 1.210 & 0.351  & &1.161 &0.269 \\
1580& 6900 & 2.027 & 1.139 & 0.342  & &1.160 &0.269 \\
1670& 7050 & 1.938 & 1.045 & 0.351  & &1.157 &0.268 \\
1775& 7250 & 1.906 & 1.035 & 0.356  & &1.152 &0.268 \\
1860& 7450 & 1.886 & 1.086 & 0.362  & &1.142 &0.267 \\
1915& 7650 & 1.924 & 1.155 & 0.357  & &1.128 &0.266 \\
1980& 8050 & 1.928 & 1.041 & 0.328  & &1.096 &0.263 \\
2017& 8400 & 1.723 & 0.975 & 0.335  & &1.064 &0.261 \\
2043& 8700 & 1.722 & 1.038 & 0.345  & &1.037 &0.258 \\
2062& 8950 & 1.773 & 1.052 & 0.344  & &1.015 &0.256 \\
2075& 9200 & 1.821 & 0.966 & 0.322  & &0.944 &0.254 \\
2087& 9450 & 1.778 & 0.894 & 0.309  & &0.988 &0.252 \\
2110& 9900 & 1.732 & 0.971 & 0.321  & &0.988 &0.248 \\
2140& 10550& 1.923 & 0.956 & 0.326  & &0.999 &0.243 \\
2168& 11150& 1.834 & 1.035 & 0.298  & &1.018 &0.238 \\
2190& 12000& 1.820 & 1.014 & 0.311  & &1.039 &0.232 \\
2199& 13000& 1.711 & 0.976 & 0.301  & &1.042 &0.226 \\
2200& 13500& 1.634 & 1.004 & 0.290  & &1.041 &0.223 \\
   \hline
\end{tabular}
%\end{minipage}
\end{table*}

%\section[]{Integral form of equations}

\end{document}